\documentclass[aps,twocolumn,superscriptaddress]{revtex4}
\usepackage{dcolumn}
\usepackage{graphicx}
\usepackage{amsmath}
\usepackage{amsfonts}
\usepackage{amssymb}
\usepackage{psfrag}
\usepackage{wrapfig}
\usepackage{subfigure}
\usepackage{makeidx}
\usepackage{bm}
\usepackage{epsf}
\usepackage{multirow}
\usepackage{upgreek}
\usepackage[colorlinks,urlcolor=blue,citecolor=blue]{hyperref}
\DeclareMathOperator{\sech}{sech}

\begin{document}
\title{Controlling dark solitons on the healing length scale}
\author{Ling-Zheng Meng}
\affiliation{School of Physics, Northwest University, Xi’an, 710127, China}

\author{Li-Chen Zhao}
\email{zhaolichen3@nwu.edu.cn}
\affiliation{School of Physics, Northwest University, Xi’an, 710127, China}
\affiliation{NSFC-SPTP Peng Huanwu Center for Fundamental Theory, Xi'an 710127, China}
\affiliation{Shaanxi Key Laboratory for Theoretical Physics Frontiers, Xi'an 710127, China}

\author{Th.~Busch}
\email{thomas.busch@oist.jp}
\affiliation{Quantum Systems Unit, OIST Graduate University, Onna, Okinawa 904-0495, Japan}

\author{Yongping Zhang}
\email{yongping11@t.shu.edu.cn}
\affiliation{Department of Physics, Shanghai University, Shanghai 200444, China}

\date{\today}
\begin{abstract}
While usually the optical diffraction limit is setting a limit for the lengthscales on which a typical alkali Bose-Einstein condensate can be controlled, we show that in certain situations control via matter waves can achieve smaller resolutions.  For this we consider a small number of impurity atoms which are trapped inside the density dip of a dark soliton state and show that any grey soliton can be obtained by just driving the impurity atoms.  By controlling the driving force on the impurity, one can therefore fully control the position and velocity of the dark soliton, and also study controlled collisions between these non-linear objects.

\end{abstract}

\maketitle

\section{Introduction}
Dark solitons in atomic Bose-Einstein condensates (BECs) are highly localised structures where the phase of the wavefunction changes by $\pi$ over a very short distance \cite{Tsuzuki,Kivshar1,Burger1,Denschlag}.  For this to be possible, the density needs to vanish across this jump, which leads to the appearance of a localised density dip on the scale of the healing length. Dark solitons are part of a larger family of structures, so called grey-solitons, where the phase jump is smaller than $\pi$ and also happens over a wider region. Grey solitons therefore do not have a point of vanishing density and they are also not stationary objects \cite{Busch1,Pelinovsky1,Fedichev,Huang1,Parker1,Parker2}. In fact, dark and grey solitons are known to be dynamically unstable and develop into accelerating  solitons, which vanish when the speed of sound is reached.

The dynamics of dark solitons is an interesting topic and over the last decades many works have studied their dynamics \cite{Busch1,Konotop2,Frantzeskakis,Becker} as a function of the external and internal parameters of the condensate supporting them. Of particular interest have been their scattering properties, however, controlling the dynamics of dark solitons is a difficult problem, since the healing length is usually much smaller than optical length scales. Optical potentials can therefore not be used for local control \cite{Becker,Carr,Carretero,Hans}, which makes independently controlling two solitons at close distance very difficult.

In this work we suggest that control of dark solitons can be achieved by using matter wave potentials instead of optical ones. These can be of significant shorter wavelength and through scattering directly effect the condensate itself. The basic idea relies on considering a small number of impurity atoms immersed into the condensate, which interact repulsively with the BEC. Assuming a density-density interaction, these atoms will be trapped inside the dark solitons, where the condensate density is low, and if their number is small, the soliton shape will be mostly unaffected. Using species-selective external potentials, one can now control the impurities, which, through their interaction with the condensate, will in turn allow one to control the soliton.

\section{Controllable motion of dark solitons}
We consider a quasi-one dimensional BEC that can be described in the mean-field limit by the Gross-Pitaevskii equation. Such a situation can be achieved by tightly confining a gas in two directions using a harmonic oscillator of frequency $\omega_\perp$, such that the dynamics in these directions is frozen out \cite{Olshanii}.
The BEC is coupled to a small number of impurity atoms and in the limit of weak interactions the set of equations describing the dynamics of the coupled  system  is given by \cite{SI1,SI2,SI3}
\begin{align}
    \label{eq:BECEOMs}
    {\rm i} \frac{\partial \psi_c}{\partial t} =& -\frac{1}{2}\frac{\partial^2\psi_c}{\partial x^2}+ |\psi_c|^2\psi_c + |\psi_i|^2\psi_c +V_c\psi_c, \\
     \label{eq:IMPEOMs}
    {\rm i} \frac{\partial \psi_i}{\partial t} =& -\frac{1}{2}\frac{\partial^2 \psi_i}{\partial x^2}+ |\psi_c|^2\psi_i +V_i\psi_i -Fx \psi_i.
\end{align}
Here $\psi_c$ and $\psi_i$ denote the wave functions of condensate and the impurity atoms, respectively. To make the above equations dimensionless, we have assumed that the mass $m$ of the atoms in the BEC and in the impurity is the same and have scaled all lengths in units of the healing length $\xi=\hbar/\sqrt{ m U n_B}$, all energies in units of $E=\hbar^2/m \xi^2=Un_B$ and time in units of $T=m\xi^2/\hbar=\hbar/Un_B$. Here the reduced quasi-one-dimensional interaction coefficient is given by $U=2\hbar a \omega_\bot$, where $a$ is the s-wave scattering length.  The density of the BEC, and therefore the background density of the dark soliton state, is given by $n_B$ and the wave functions $\psi_c$ and $\psi_i$ are renormalized to this density.  Using typical experimental parameters for a BEC experiment of $^{87}\textrm{Rb}$ atoms we have $a=100a_B$ \cite{vanKempen,Widera} with $a_B$ being the Bohr radius, $\omega_\bot/2\pi=500{\rm Hz}$, and $n_B=10^9$ per meter. The scales of units then become $\xi=0.15\upmu {\rm m}$ and $T=0.03{\rm ms}$. Since the impurity component only contains a very small number of atoms, we neglect a potential nonlinear interaction within this component in Eq.~(\ref{eq:IMPEOMs}), however we assume that a selective force $F$ is applied to the impurity atoms only. This is described by the last term in the Eq.~(\ref{eq:IMPEOMs}). The external potentials that the BEC and impurity feel are given by  $V_c$ and $V_i$, respectively.

\begin{figure}[tb]
\includegraphics[width=\linewidth]{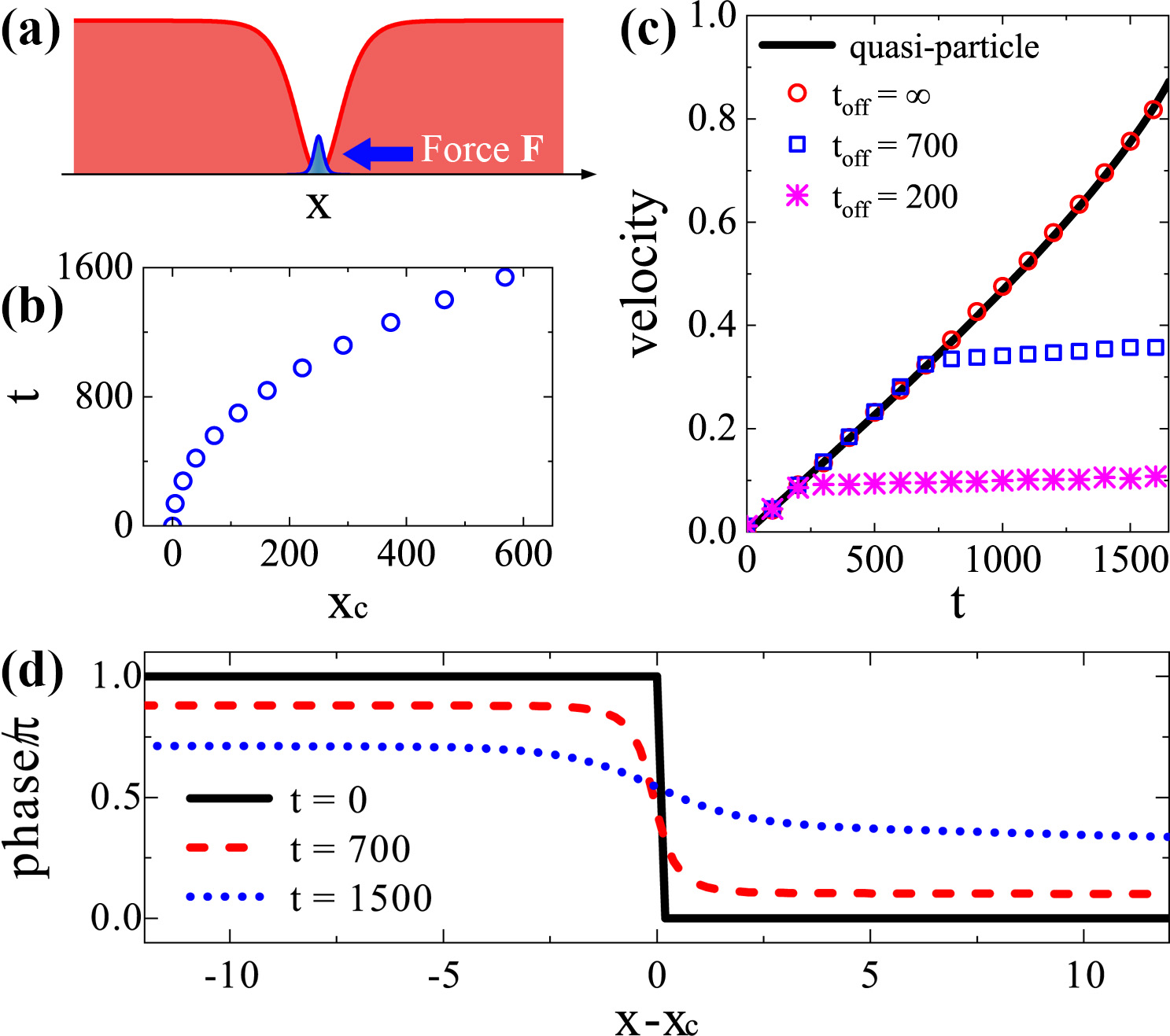}
\caption{(a) Schematic of a dark soliton with few impurity atoms trapped in the density dip. The external force $F$ is applied only to the impurity atoms. (b) Time evolved center of the soliton density dip with the state at $t=0$ being a dark soliton. (c) Velocity of the accelerated soliton as a function of time. The red circles, blue squares, and purple asterisks are obtained by switching off the linear external potential at different times and the black solid line is given by Eq.~\eqref{eq:EqOfMotion}. (d)  Snapshots of the phase of the condensate during the acceleration process, compared to the initial state. The parameters are $\varepsilon = 0.3$ and $F=-0.01$.  }\label{fig:Schematic}
\end{figure}

We first study the effect of an external force on the impurity for the case of $V_c=V_i=0$. If the initial state of the condensate is given by the exact dark soliton solution, $\psi_c = \tanh(x)$, then  the ground state of the impurity can be approximated as the ground state of the mean-field matter wave potential [see schematic in Fig.~\ref{fig:Schematic}(a)], $
 \psi_i = \varepsilon \sech(x)$.
Here the normalisation is chosen such that $\varepsilon^2 \ll 1$ to ensure that the number of impurity atoms is much smaller than the number of atoms in the BEC.

While one would naively expect that the dark soliton will be dragged in the direction of the force by the accelerating impurity atoms, one can see from the numerical evolution of the GP equations in Fig.~\ref{fig:Schematic}(b) that the joint system moves in the direction opposite to the force. This might seem counterintuitive at first, but it can be easily understood by the concept of negative mass which is associated with dark and grey solitons \cite{Busch1,Konotop2,Busch2,Aycock}. We use a variational approach to get the equation of motion for the center of the soliton, from which the negative mass effect can been seen. For this we introduce trial wavefunctions as \cite{Kivshar:95}
\begin{align}\label{trialfunction}
\psi_c(t) &= \textrm{i}\sqrt{1-p^2(t)} + p(t)\tanh\left[\frac{x-x_c(t)}{w(t)}\right], \\
\psi_i(t) &= \varepsilon(t) \sech\left[\frac{x-x_c(t)}{w(t)}\right] \textrm{e}^{\textrm{i} \{\phi_0(t) + [x-x_c(t)]\phi_1(t)\}},
\end{align}
where $p(t)$ accounts for the depth of the dark soliton, $w(t)$ for its width, and $x_c(t)$ for its position. The amplitude of the impurity is described by $\varepsilon(t)$. For the initial static dark soliton the parameters are $p(0)=1, w(0)=1, \varepsilon(0)=\varepsilon$ and $\dot{x}_c(0)=0$ and the evolution is guided by the Lagrangian of the joint system
\begin{align}\label{eq:Lagrangian}
\mathcal{L}=&\int_{-\infty}^{+\infty}  \bigg[ \frac{\textrm{i}}{2}(\psi_c^* \partial_t \psi_c - \psi_c \partial_t \psi_c^*)(1-\frac{1}{|\psi_c|^2})- \frac{1}{2}|\partial_x \psi_c|^2 \nonumber \\
&
+ \frac{\textrm{i}}{2}(\psi_i^* \partial_t \psi_i - \psi_i \partial_t \psi_i^*)  - \frac{1}{2}|\partial_x \psi_i|^2 + Fx|\psi_i|^2 \nonumber \\
&- \frac{1}{2}(|\psi_c|^2-1)^2 - |\psi_i|^2(|\psi_c|^2-1) \bigg] \textrm{d}x.
\end{align}
Substituting the trial wavefunctions into the Lagrangian and applying the Euler-Lagrangian equations $\frac{\textrm{d}}{\textrm{d}t}(\frac{\partial \mathcal{L}(t)}{\partial\dot{\alpha}}) = \frac{\partial \mathcal{L}(t)}{\partial\alpha}$
with  $\alpha=\{ w(t), p(t), x_c(t)\}$ and $\dot{\alpha}$
denoting $\frac{\textrm{d} \alpha}{\textrm{d}t}$, we derive the coupled equations for the motions of $\alpha$.
Since for the system we consider the impurity to be the minority component ($\varepsilon^2(t)\ll1$), almost all terms that contain $\varepsilon^2(t)$ can be neglected. The exception is the term describing the potential energy, which can be seen to increase with $x_c(t)$ and which can therefore become comparable to the energy of the dark soliton. In fact, under the local density approximation, the  potential energy can be evaluated as $E_p = \int_{-\infty}^{+\infty} -F x |\psi_i|^2 \textrm{d}x \approx -F x_c \int_{- \infty}^{+\infty} |\psi_i|^2 \textrm{d}x = - F N_i  x_c(t)$, where $N_i= 2\varepsilon^2(t)w(t) = 2\varepsilon^2$ is the particle number of the impurity component. The effective force applied to the soliton is therefore $F_{\textrm{eff}} = -\nabla E_p = F N_i$.
At last, the equation of motion for the center can be simplified as
\begin{equation}
    \label{eq:EqOfMotion}
    \ddot{x}_c = -\frac{F_{\textrm{eff}}}{4\sqrt{1-\dot{x}_c^2}}.
\end{equation}
This equation shows that the center of mass motion behaves as a quasi-particle with the negative mass $m_\mathrm{eff}=-4\sqrt{1-\dot{x}_c^2}$ and it clearly shows that the soliton moves against the force applied to it. In Fig.~\ref{fig:Schematic}(c) we show the velocity of the dark soliton calculated from the numerical evolution of the GP equations and from the quasi-particle equation in Eq.~(\ref{eq:EqOfMotion}). The two results match very well, which also indicates that the trial wavefunctions is a good description of  the system. We also show the phase of the BEC component around the center of the dark soliton dip in Fig.~\ref{fig:Schematic}(d), where one can see that  for the different acceleration times the resulting condensate states have the appropriate phase changes expected for grey solitons. The results shown in Figs.~\ref{fig:Schematic}(c) and \ref{fig:Schematic}(d) therefore indicate that during the evolution the background of the soliton is not significantly perturbed, which suggests that a dark soliton can be accelerated by applying a force to an impurity.  In that sense, the state of large condensate system is controlled precisely by driving the small matter wave.

This can also be seen by deriving the equation of motion for the center of the soliton directly from the dispersion relation of the exact scalar dark soliton solution, while ignoring the kinetic energy of impurity atoms and the nonlinear coupling between the impurity and condensate. In this case the total energy of the system is given by the excitation energy of the dark soliton \cite{Kivshar:95}, i.e., $E_s= \frac{4}{3}(1-v^2)^{3/2}$. With the potential energy given by $E_p  \approx -  F N_i x_c$, the trajectory of the dark soliton can be derived based on the conservation of energy $E_s + E_p = E_s\big|_{v=v_0}$, where we assume that the initial center position of the soliton is at $x_c|_{t=0}=0$ \cite{Zhao,Meng}. This approach gives the same equation of motion as the variational one in Eq.~\eqref{eq:EqOfMotion}, which furthermore indicates that the dark soliton background is kept nearly perfect during the acceleration process. It should therefore be possible to control the soliton velocity precisely and without exciting  dispersive waves or generate excitations on the background \cite{Becker,Carr,Carretero,Radouani2,Burger2,Engels}.

\begin{figure}[b]
\begin{center}
\includegraphics[width=80mm]{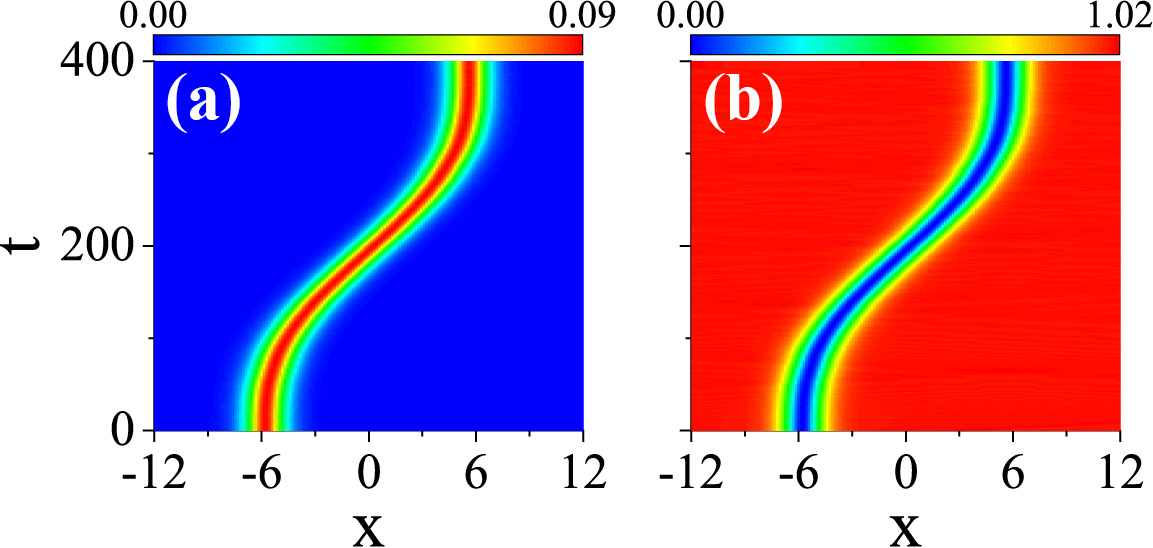}
\end{center}
\caption{The evolution of the (a) impurity component and (b) condensate density. A force $F = -0.01 \sin(\pi t/200)$ is applied to the impurity, which changes sign at $t=200$.
Here $\varepsilon = 0.3$. }\label{fig:AccDec}
\end{figure}

In addition to accelerating the soliton states to different velocities through the impurities, it also possible to slow them down and, in fact, return them to their initial dark state. We show an example of this in Fig.~\ref{fig:AccDec}, where the impurity component coupled to an initially stationary dark soliton  experiences a slowly varying force of the form $F = -0.01 \sin(\pi t/200)$. One can see that the dark soliton is accelerated until $t=200$ and then begins to decelerate and return to be the initially static state at $t=400$.  In order to quantify how close the final state is to a translated dark soliton, we define a fidelity
\begin{equation}
    \mathcal{I}=\frac{\left|\int^{L}_{-L} \psi_v^* \psi_{D} \textrm{dx}\right|^2}{\left|\int^{L}_{-L} \psi_v^* \psi_v \textrm{dx}\right|^2},
\end{equation}
where $\psi_v$ is the target state given by the exact scalar dark soliton solution, $\psi_{D}$ is the wavefunction at the end of the acceleration and de-acceleration process, and $2L$ is a finite area around the soliton center within which the density has well returned to the bulk value.  The fidelity $\mathcal{I}$ takes value from $0$ to $1$, and a larger $\mathcal{I}$ indicates that the created state is more consistent with the exact solution. For the example shown in Fig.~\ref{fig:AccDec} the fidelity at $t=400$ within a region given by $L=16$ is given by $\mathcal{I} = 0.9975$. This shows that not only the region around the density dip is close to the exact dark soliton wavefunction, but also that no significant sound waves were emitted during the whole dynamical process \cite{Parker1,Parker2,Parker3,Radouani1,Pelinovsky2}. This fact is of particular importance for the development of techniques that allow for highly accurate control over localised structures in condensates \cite{Carr,Carretero,Hans,Radouani2,Radouani1,Konotop1,Hakim,CRaman,Bongs,Pavloff,Theocharis1,Theocharis2,Lyu,Fritsch}.

\begin{figure}[t]
\begin{center}
\includegraphics[width=65mm]{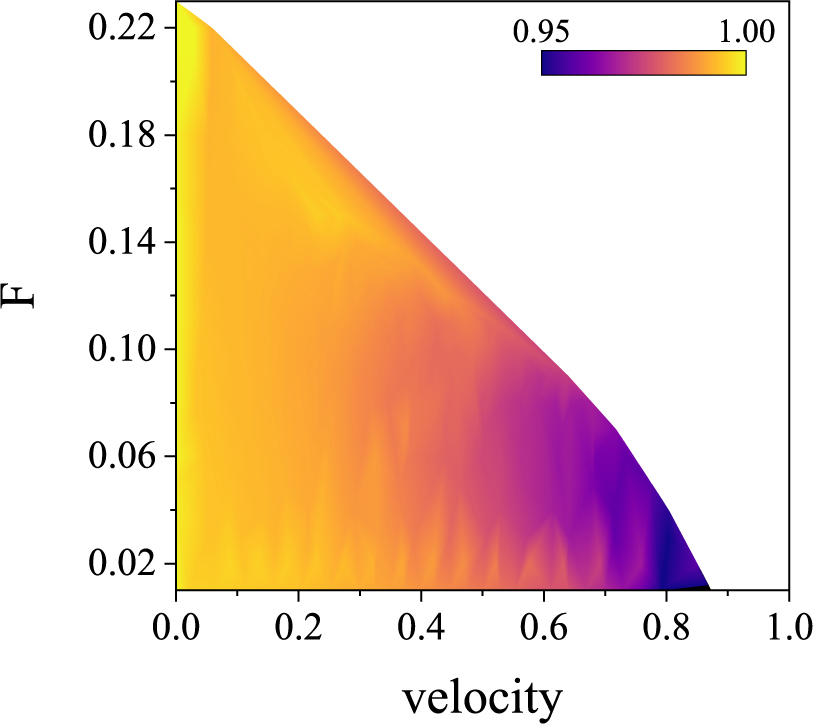}
\end{center}
\caption{Fidelities of grey solitons generated by applying different strengths acceleration potentials $-F x$ to the impurity component. The blank area represent parameter regimes in which the force is too strong and the impurity atoms are pulled out of the traps formed by the condensate density dips.  Here $\varepsilon = 0.3$.}\label{figfidelity}
\end{figure}

Being able to control the motion of dark solitons also can be used to create grey solitons with high precision.
In Fig.~\ref{fig:Schematic}(c) we show results for two cases of accelerating dark solitons and suddenly switching off the force acting on the impurity at a finite time. One can see that it is not only possible to create grey solitons of specific velocities this way, but also that these structures move with essentially constant velocities for longer times. In Fig.~\ref{figfidelity}, we show the fidelity of grey solitons generated in this way as a function of the applied force and resulting velocity of grey solitons. While a too strong acceleration of the impurity results in it leaving the dark soliton density dip (white areas in Fig.~\ref{figfidelity}), for all parameters where the trapping inside the dark soliton remains one can see that the fidelities are $\mathcal{I}\geq0.95$. Furthermore, fewer impurity atoms and weaker external force can be used to broaden the speed regime of grey solitons, but much longer acceleration (and numerical evolution) times will be needed.

\begin{figure}[t]
\begin{center}
\includegraphics[width=85mm]{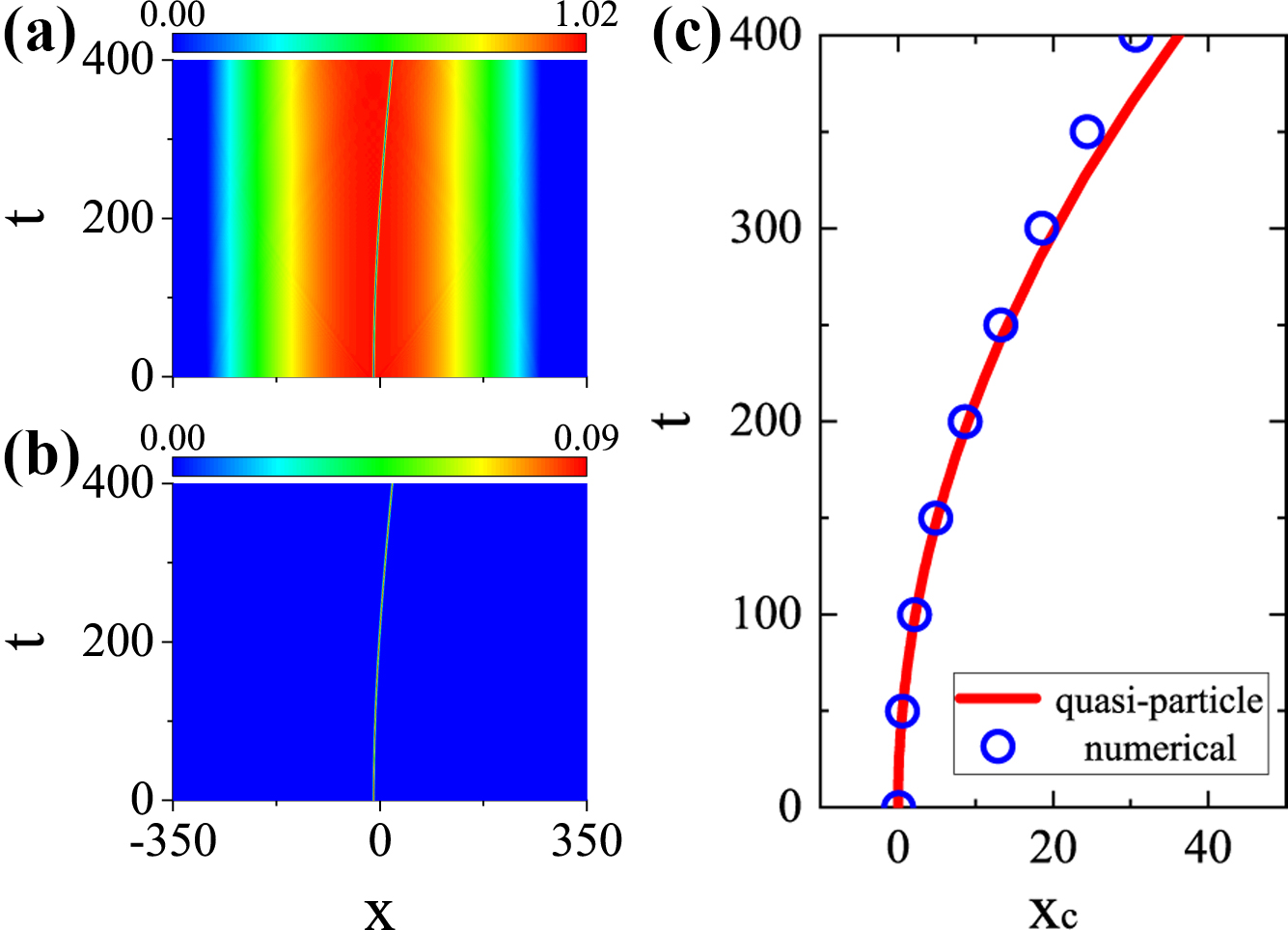}
\end{center}
\caption{Densities of the (a) condensate dark soliton and (b) impurity component in the presence of a weak harmonic trap, $\eta=0.005$, under a force on the impurity. (c) Position of the minimum of the dark soliton dip from the numerical evolution (circles)  compared to the quasi-particle motion given by Eq.~(\ref{eq:EqOfMotion}) (full line). }\label{fig:Har1}
\end{figure}

\section{The effect of the external Harmonic traps}

In real BEC experiments external traps are inevitably involved to confine the condensate atoms. While box potentials can be realised experimentally, we study in the following the effect of the much more common harmonic traps on the controllable motions of dark solitons. For simplicity we assume that the harmonic trapping strengths seen by the BEC and the impurity components are the same,  $V_c=V_i=\frac{1}{2}\eta^2 x^2$, where we have defined the dimensionless quantity $\eta=m\omega_x \xi^2/\hbar$ with $\omega_x$ being the trapping frequency in the axial direction. We first study a weak harmonic trap, $\eta=0.005$, which corresponds to $\omega_x/2\pi=25{\rm Hz}$, and prepare the initial dark soliton within a background of Thomas-Fermi shape. The initial state of the impurity can be chosen the same as the case without a trap, as it only sees the mean-field potential of the condensate dark soliton over its healing length extension. The evolution of this initial state when a force $F=0.01$ is applied to the impurity component is shown in Fig.~\ref{fig:Har1}. As before, the force accelerates the dark soliton and the impurity to move together in the direction opposite to the force. During the acceleration, the background stays essentially unperturbed and no phononic excitations in the BEC component are visible. Since the density of the background condensate changes slowly on the healing lengths scale, the dynamics of the center of the dark soliton still matches very well with the expected trajectory in the free case given by Eq.~(\ref{eq:EqOfMotion}) and for short acceleration times the two trajectories coincide [see panel (c)]. However, it can also be seen that the change in the background density leads to a slowing down of the acceleration over time, as one would expect from a freely oscillating dark-bright soliton as well \cite{Busch2}.

\begin{figure}[t]
\begin{center}
\includegraphics[width=85mm]{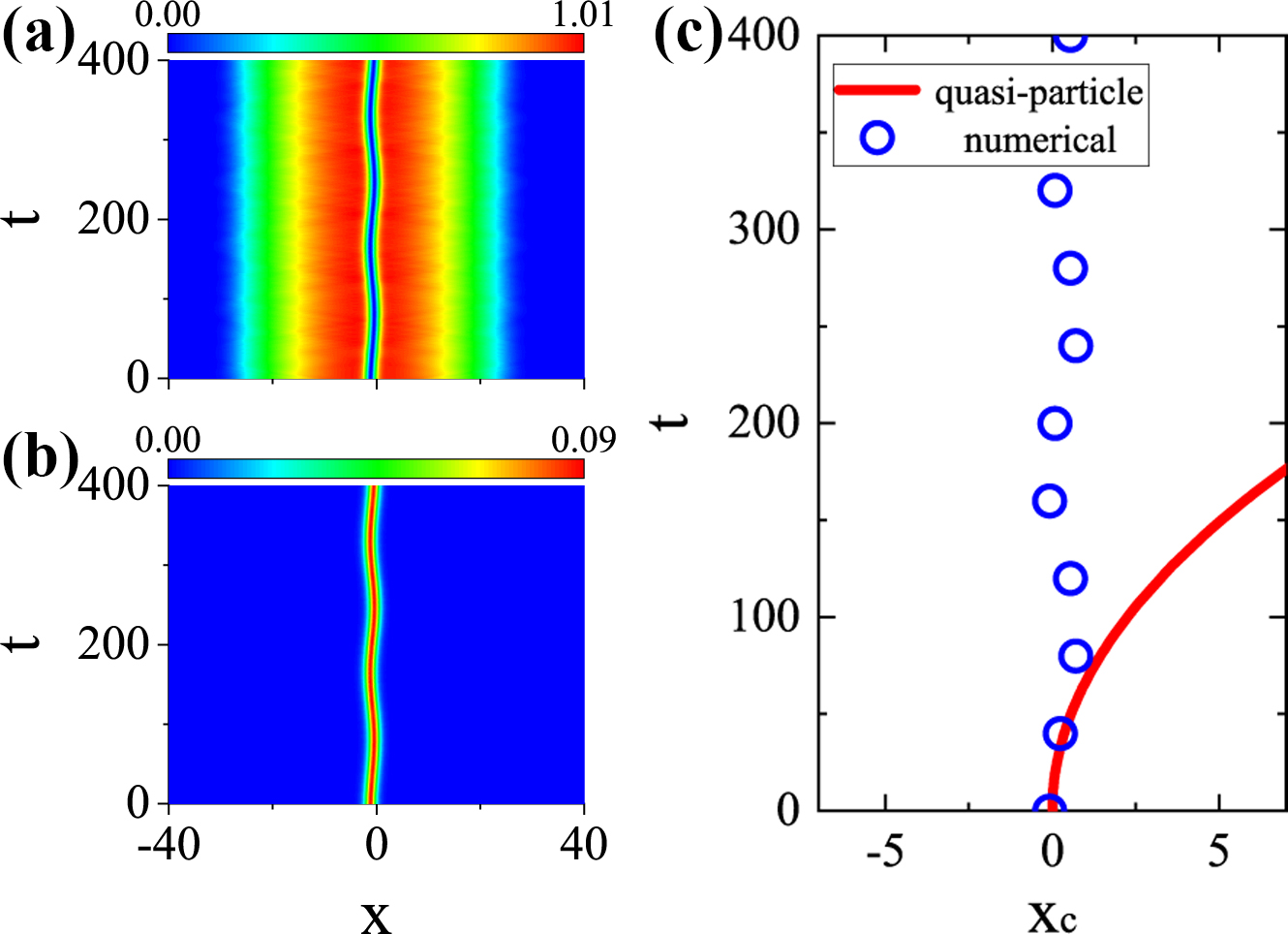}
\end{center}
\caption{Same as these in Fig.~\ref{fig:Har1}, but in the presence of a stronger harmonic trap with $\eta=0.05$. } \label{fig:Har2}
\end{figure}

It is interesting to wonder if the slowdown of the acceleration due to the inhomogeneous background density in the harmonic trap can completely stop the joint system. To explore this we study a system that is more strongly trapped, $\omega_x/2\pi=250{\rm Hz}$ and $\eta=0.05$, in which the inhomogeneity happens over shorter length scales. For the same, weak strengths of the accelerating force, $F=-0.01$, the resulting dynamics is shown in Fig.~\ref{fig:Har2} and one can see that the trap indeed stops the acceleration. However, no stationary state is obtained, but instead the joint system starts to oscillate around the traps center.
This can be understood by realising that the number of impurity atoms is small and the strength provided by nonlinear coupling between them and the condensate is limited. When the force exerted by harmonic potential becomes stronger than the weak constant force on the impurity, the well known dynamics of a dark soliton oscillating in a harmonic trap becomes dominant \cite{Busch1}. Therefore, the predictions of Eq.~\eqref{eq:EqOfMotion} are only valid when the harmonic trap is weak, whereas when the condensate is strongly trapped, the oscillatory dynamics is dominant and the constant force only causes a small position shift of the oscillation center.

\begin{figure}[tb]
    \includegraphics[width=85mm]{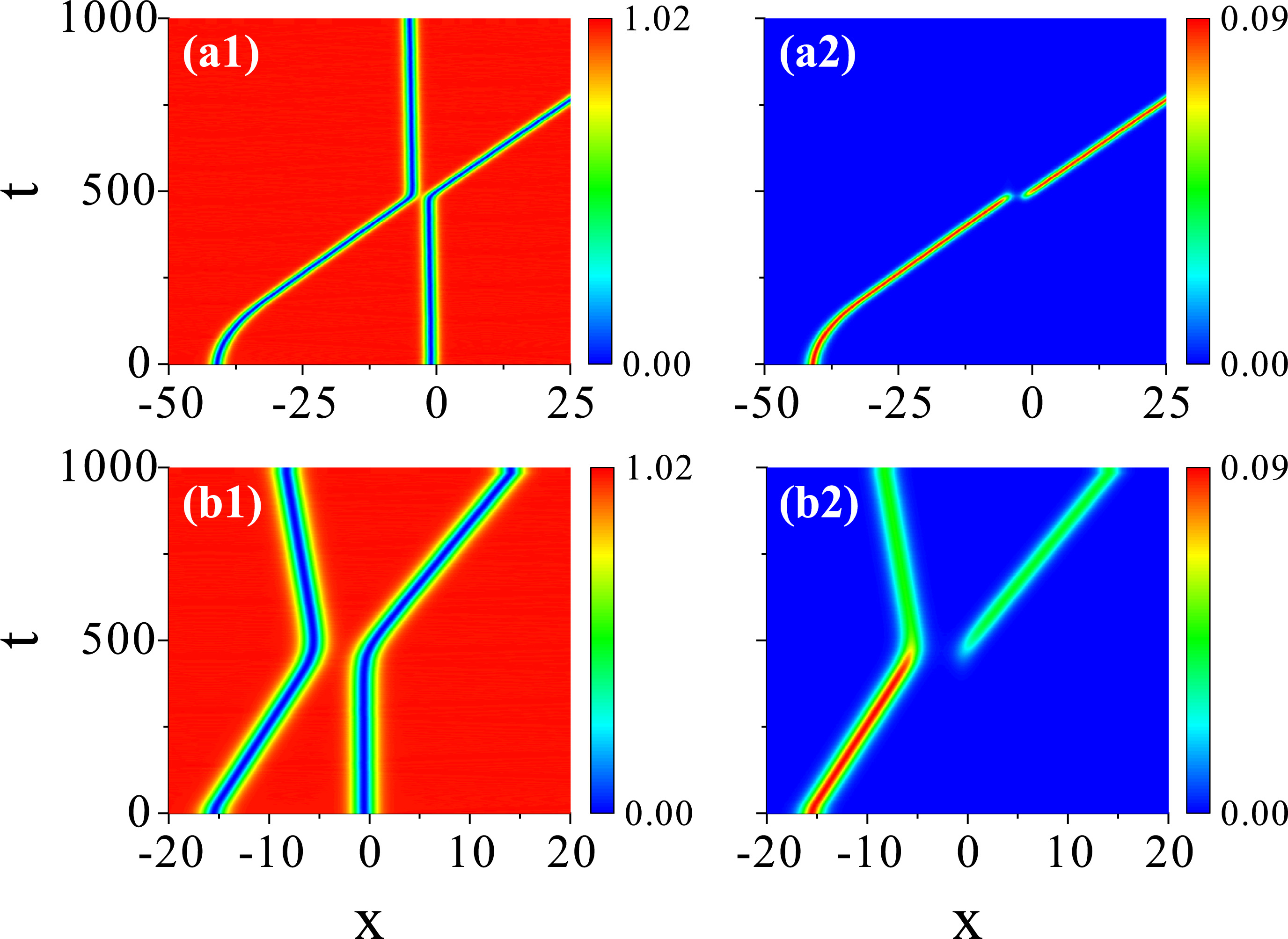}
    \caption{Densities of the two components for (a) an elastic and (b) an inelastic  collision between two dark solitons. The soliton initially on the left hand side is accelerated by means of driving the impurity, and the external force is switched off at (a) $t=200$ for the elastic case and (b) $t=50$ for the inelastic case. The soliton initially on the right is impurity free. The parameters are $\varepsilon = 0.3$ and $F=-0.01$.}
    \label{fig:collision}
\end{figure}

\section{Applications to dark solitons collision}

Interactions between dark solitons in BECs are an extensive research area which has produced many beautiful theoretical and experimental results \cite{Huang,Burger2,Weller,Chang,Stellmer,Theocharis3}.
Yet, controlling such collisions to a high degree is quite difficult, with one challenge being the precise acceleration of the dark solitons without dynamically affecting the entire BEC system.
Two ways previously experimentally used for creating grey solitons to study interactions between them are matter-wave interference  \cite{Weller,Chang,Theocharis3} and phase imprinting and density engineering techniques \cite{Burger2,Stellmer}. However, grey solitons generated by these method are often accompanied by a significant number of sounds waves, which makes precise control of their velocity difficult. Creating a stationary dark solitons can therefore be thought of as a much more well defined problem, with the acceleration done in a next step through the method discussed here. In particular, the sound generated when preparing the dark soliton could be allowed to decay before the soliton gets accelerated.

We therefore investigate a scheme to realize a collision between two dark solitons by initially preparing two static dark solitons away from each other and adding a small number of impurity atoms to one of them. Applying a force to the impurity component for a certain amount of time will accelerate the dark soliton to the required speed, and have it then collide with the second soliton subsequently. Two examples of such a protocol are shown in Fig.~\ref{fig:collision}, where the soliton on the left is accelerated and collides with the one  on the right. If the collisional speed is large (panels (a1) and (a2)) one can see that the impurity component maintains its momentum and the collision can be thought of as elastic. This is also indicative of an integrable model. For collisions at lower velocities (panels (b1) and (b2)) the impurity component splits between the two dark solitons, and the interaction can be thought of as inelastic. In general the switch-over between elastic and inelastic collisions depends on the number of impurity atoms present. As before, in either scattering regime the BEC background is not affected and this method therefore allows one to carefully study very clean soliton-soliton collisions \cite{Zakharov}.

\section{Conclusion}

We have shown that the dark soliton state of a large condensate system can be controlled well by a small matter wave potential that is selectively accelerated or decelerated. Since the impurity is trapped inside the dark soliton density dip, it allows to control the condensate through a force applied over the healing length scale, which can be much shorter than a typical optical wavelength.

In particular, this allows to turn a stationary dark soliton state into a moving grey soliton one with high precision and without exciting any significant phonon modes within the background condensate density. This process can also be reversed or used to study collisions between solitons in a highly controlled way.
While the creation of stationary dark solitons is still an experimentally difficult process, decoupling it from the acceleration process for creating grey solitons is a significant simplification for studying soliton dynamics in a broad range of settings.

Our work above lays out the foundations of this new idea and demonstrates some of the advantages, for example the possibility to accelerate excitation free. Further studies to explore the limits and also higher dimensional settings are the important next steps.

\section*{Acknowledges}
L.-C. Z. was supported by National Natural Science Foundation of China with Grants No. 12022513 and
 12235007.  Y.Z. was supported by National Natural Science Foundation of China with Grants No.11974235 and
11774219. This work was supported by the Okinawa Institute of Science and Technology Graduate University. Y.Z.and T.B. were also supported by the Japan-China Scientific Cooperation Program between JSPS and NSFC under Nos. 120227414 and 12211540002.

\end{document}